\documentstyle[12pt,a4]{article}
\begin{document}
\hyphenation{anti-fermion}
\topmargin= -20mm
\textheight= 230mm
\baselineskip = 0.33 in
 \begin{center}
\begin{Large}

{\bf  g-factor of a tightly bound electron   }

\end{Large}

\vspace{2cm}

T. ASAGA\footnote{e-mail: tasaga@shotgun.phys.cst.nihon-u.ac.jp}, 
T. FUJITA\footnote{e-mail: fffujita@phys.cst.nihon-u.ac.jp} and 
M. HIRAMOTO\footnote{e-mail: hiramoto@phys.cst.nihon-u.ac.jp}

Department of Physics, Faculty of Science and Technology  
  
Nihon University, Tokyo, Japan

\vspace{3cm}

{\large ABSTRACT} 

\end{center}

We study the hyperfine splitting of an electron in hydrogen-like \ 
$^{209}Bi ^{82+} $. It is found  that the hfs energy splitting can be 
 explained well by considering the g-factor reduction  
due to the binding effect of a bound electron. 
We determine for the first time the experimental value of the magnetic 
moment of a tightly bound electron. 

\vspace{2cm}

PACS numbers:  31.30Jv, 31.30Gs 

\vspace{1cm}

( Phys. Rev. {\bf A} in press )

\newpage

Recently, the hyperfine structure (hfs) of  electronic atoms has received 
a renewed interest [1-7]. Experimentally, this situation 
may well be due to recent technical 
progresses  which enable to carry out
precision measurements of the hyperfine splitting. 
Theoretically, the accurate determination of the hfs splitting 
 is expected to allow a novel test of QED 
corrections under the strong magnetic field. 

For the investigation of the hyperfine structure in electronic atoms, 
it is always ideal if one makes hydrogen-like atoms. 
Indeed, this is done by Klaft et al.[1] who made a very nice 
measurement of the hyperfine energy splitting for a tightly 
bound electron state in hydrogen-like  $^{209}Bi^{82+} $. 
They measured the wave length $\lambda$ of the $M1$ transition 
between $F=4$ and $F=5$ hyperfine 
levels of the ground state of  hydrogen-like \ 
$^{209}Bi ^{82+} $, and obtained $\lambda_{exp} = 243.87(4) $ nm. 
This value should be compared to the wave length calculated with 
a point Coulomb interaction. A theoretical value is  
$\lambda_{point}=212.7$ nm for hydrogen-like $^{209}Bi^{82+} $.  
(Note that this number is obtained with the free electron g-factor.) 

This difference between hfs level splitting
 has been discussed by Finkbeiner et al.[2] and  
by Schneider et al.[3].  Finkbeiner calculated the hyperfine anomaly 
due to the finite size of the nucleus. They showed that the observed 
value of the hfs splitting can be understood by varying the magnetization 
distribution of the nucleus. However, Schneider et al. showed that 
the hyperfine anomaly of the electronic atom in $^{209}Bi^{82+} $ 
is much smaller than the observed difference 
$\delta \lambda = \lambda_{exp}-\lambda_{point} $.  
 Their calculated value of Bohr-Weisskopf effect [8] is  
$ \delta \lambda_{BW} \approx 3.5 \ {\rm nm} $, as compared 
to the observed value of $\delta \lambda =31.17 \ {\rm nm} $. 
Indeed, one can easily confirm 
oneself that the hfs anomaly $\epsilon$ due to Bohr-Weisskopf 
effect is not much different from $1 \  \%$ if 
one makes use of the formula given in ref.[9]. This is obviously 
much too small compared to $\delta \lambda /\lambda_{point} = 0.149$. 
Schneider et al.[4] also estimate  other QED corrections such as 
vacuum polarization. They find 
$ \delta \lambda_{vp} = -1.6 \ {\rm nm}  $, which is also small. 
Thus, up to now, all of the theoretical predictions 
 are much too small to explain the observed difference 
$\delta \lambda$.

In this Letter, we show that the observed difference $\delta \lambda$ 
can be explained well by considering the g-factor reduction of the bound 
electron in $^{209}Bi^{82+} $. This can easily be seen since the change 
of the g-factor of a bound electron 
 due to the binding effect can be written for a point charge case as [10-11] 
$$ g_{BC}= -{2\over 3}(1-\sqrt{1-\gamma^2} ) g_0 \eqno{(1)} $$
where $\gamma$ is defined as $\gamma ={Ze^2\over{\hbar c}}$. 
$g_0$ denotes the g-factor of a free electron. For $^{209}Bi^{82+} $, 
we find 
$$ g_{BC}= -0.136 \  g_0  . \eqno{(2)} $$
Therefore, the wave length $\lambda$ should be corrected as 
$$ \lambda = {g_0\over{(g_0+g_{BC})}}\lambda_{point} .  \eqno{(3)} $$
This gives  
$$\lambda = 246.2 \ \  {\rm nm} \eqno{(4)} $$ 
which should be compared to the observed 
value of $\lambda_{exp} = 243.87(4) $ nm. The agreement between 
theory and experiment is now remarkably good. Thus, the hfs splitting 
in electronic atom with high $Z$ is understood by the g-factor 
reduction due to the binding effect of the electron in $^{209}Bi^{82+} $. 

The reduction of the g-factor in muonic atoms has been already 
measured by Yamazaki et al.[12]. 
 The observed reduction of a bound muon 
for muonic $^{208} Pb$ atom is 
$$ g_{BC}^{exp}({\rm muon}) = ( -0.047 \pm 0.022 ) g_0 . \eqno{(5)} $$ 
This is compared to the theoretical g-factor reduction in muonic  
$^{208} Pb$ atom [11] 
$$g_{BC} (muon) = -0.032 \ g_0 \eqno{(6)} $$ 
which is consistent with the observed g-factor reduction. 
In muonic atoms, 
the g-factor reduction is much smaller than that of the electronic 
atoms since the $1s_{1\over 2}$  muonic orbit 
in $^{208} Pb$ atom is almost inside 
the nucleus and thus the Coulomb interaction is much weaker 
than the point Coulomb case, which is almost the case in electronic 
atoms. 

Now, in what follows, we turn around the argument and try to extract the 
"experimental" g-factor of a tightly bound electron 
in high Z electronic atoms. 
For this purpose, we calculate the hfs anomaly of the electronic atoms 
in $^{209}Bi^{82+} $. All the necessary formula are given in ref.[9]. 
Here, we only write the results of the hfs constant $a_I$ 
for the $1s_{1\over 2}$ state,
$$ a_{I}= a_{I}^{(0)} ( 1+ \epsilon  )    \eqno{(7)}  $$
where $a_I^{(0)}$ denotes the hfs constant for a point 
charge case and is written as 
$$ a_I^{(0)} = {4ge\mu_N\over{3I}} \mu \int_0^\infty 
F^{(1s)}G^{(1s)} dr  . \eqno{(8)} $$
Here, $g$ denotes the g-factor of an electron and $\mu$ is the magnetic 
moment of the nucleus.  $F^{(1s)}$ and $G^{(1s)}$ 
are the small and large components of the radial part of Dirac wave 
functions for the $1s_{1\over 2}$ state. \  
 $\epsilon$ can be expressed  as 
$$\epsilon =-0.62 b^{(1s)} < (R/R_0)^2> -0.38 
b^{(1s)}{< (R/R_0)^2> \over{\mu} } 
\left[\pm g_s {3(I+{1\over 2})\over{4(I+1)}}+
{3\over 4}{g_s\over{g_s -g_\ell }}(\mu-\mu_{sp} ) \right]   
  \eqno{(9)} $$
for $I= \ell \pm {1\over 2}$. 
$R_0$ is a nuclear radius and can be given as
$ R_0=r_0 A^{ 1\over 3} $ with $r_0=1.2 $ fm. 
On the other hand, $b^{(1s)}$ is a constant 
which can be calculated in terms of relativistic electron 
wave functions with finite extension of the nucleus [8], 
and we find here   $b^{(1s)}= 0.035$. $\mu_{sp}$ denotes 
the magnetic moment of the single particle state with 
$I= \ell \pm {1\over 2}$.  

The energy splitting $\Delta E$ between $F=5$ and $F=4$ hyperfine 
states is related to $a_I$ as 
$$ \Delta E = 5a_I  . \eqno{(10)} $$
The wave length $\lambda $ of the $M1$ transition is related to $\Delta E$ 
as 
$$ \lambda = {2\pi\over{\Delta E}}  . \eqno{(11)} $$
Now, we evaluate the hfs anomaly, considering 
 the effects of the core polarization with $\Delta \ell =0$ transition. 
For $^{209}Bi $, the single particle state is $ |\pi (1h_{9\over 2}) >$. 
The observed  magnetic moment 
of $^{209}Bi $ nucleus is $\mu = 4.1106 \ n.m.$ 
Also, the magnetic moment of the single particle state is 
$\mu_{sp}= 2.624 \ n.m.$
Therefore, we find  $\epsilon = -0.0083$. 
The calculated wave length shift due to the Bohr-Weisskopf effect becomes 
$$ \delta \lambda_{BW} = -\epsilon \lambda_{point} = 1.8 
\ \ {\rm nm} . \eqno{(12)} $$  
In addition, there is a contribution from higher order QED 
corrections [13,14]. This includes the vacuum polarization 
as well as radiative corrections, and  is estimated to be
$$ \delta \lambda_{QED} = -1.0 \ \ {\rm nm} . \eqno{(13)} $$
Therefore, the theoretical 
value of the wave length including the Bohr-Weisskopf effect
 and QED corrections becomes 
$$ \lambda_{theory} = \lambda_{point}+ \delta \lambda_{BW} 
 +\delta \lambda_{QED} = 213.5 \
  {\rm nm} .  \eqno{(14)} $$ 
From this value, we can extract the "observed " g-factor of 
the electron in $^{209}Bi^{82+} $ atom. 
We find 
$$ g_{BC}^{exp} =-0.125 \ g_0 . \eqno{(15)} $$
This value should be compared to the theoretical value 
of $g_{BC}=-0.136 \ g_0$. 
The extracted experimental value is quite close to the theoretical one. 
Note that eq.(1) is obtained with a point charge case. The finite 
size effect on the $g_{BC}$ may decrease  its magnitude by a few percents, 
which is consistent with the observed value of eq.(15). 

\vspace{1cm}

In summary, we have shown that the hyperfine splitting of the 
ground state of hydrogen-like $^{209}Bi^{82+} $ is explained by 
the g-factor reduction of a tightly bound electron. 
This is the first experimental determination of the g-factor 
reduction of a tightly bound electron. 

\newpage
\underline{\large References}
\vspace{0.5cm}

1. I. Klaft, S. Borneis, T. Engel, B. Frick, R. Grieser, G. Huber, 
T. K\"uhl, D. Marx, 

\quad R. Neumann, S. Schr\"oder, P. Seeling and L. V\"olker, 
Phys. Rev. Lett. {\bf 73}, 2425 (1994)

2. M. Finkbeiner, B. Frick and T. K\"uhl, Phys. Lett. 
{\bf A176}, 113 (1993)

3. S.M. Schneider, J. Schaffner, G. Soff and W. Greiner, 
J. Phys. {\bf B26}, L581 (1993) 

4. S.M. Schneider, W. Greiner and G. Soff, 
Phys. Rev. {\bf A50}, 118 (1994)

5. T. Asaga, T. Fujita and K. Ito, Z. Phys. {\bf A359}, 237 (1997)

6. K. Enders, O. Becker, L. Brand, J. Dembczynski, G. Marx, G. Revalde, 

\quad P.B. Rao and G. Werth, Phys. Rev. {\bf A52}, 4434 (1995)
  
7. M. Wada, K. Okada, H. Wang, K. Enders, F. Kurth, T. Nakamura, S. Fujitaka, 

\quad  J. Tanaka, H. Kawakami, S. Ohtani and I. Katayama, 
to appear in Nucl. Phys. {\bf A}

8. A. Bohr and V.F. Weisskopf, Phys. Rev. {\bf 77}, 94 (1950) 

9. T. Fujita and A. Arima, Nucl. Phys. {\bf A254}, 513 (1975)

10. K.W. Ford, V.W. Hughes and J.G. Wills, 
Phys. Rev. {\bf 129}, 194 (1963)

11. T. Fujita, Ph. D Thesis, UTPN-55 (1974)

12. T. Yamazaki, S. Nagamiya, O. Hashimoto, K. Nagamine, K. Nakai, 

\quad K. Sugimoto and K.M. Crowe, Phys. Lett. {\bf 53B}, 117 (1974)

13. S.J. Brodsky and G.W. Erickson, Phys. Rev. {\bf 148}, 26 (1966)

14. G.T. Bodwin and D.R. Yennie, Phys. Rev. {\bf D37}, 498 (1988)

\end{document}